\documentclass{aastex61}

\begin{document}
\title{Energy origination and triggering mechanism of a series of homologous confined flares}

\email{yanxl@ynao.ac.cn}

\author{Guorong Chen}
\affiliation{Yunnan Observatories, Chinese Academy of Sciences, Kunming 650011, China}
\affiliation{University of Chinese Academy of Sciences, Yuquan Road, Shijingshan Block Beijing 100049, China.}

\author{Xiaoli Yan}
\affiliation{Yunnan Observatories, Chinese Academy of Sciences, Kunming 650011, China}
\affiliation{Center for Astronomical Mega-Science, Chinese Academy of Sciences, 20A Datun Road, Chaoyang District, Beijing, 100012, China.}

\begin{abstract}
Using the H$\alpha$ data from the New Vacuum Solar Telescope (NVST) at the Fuxian Solar Observatory together with multi-wavelength images and magnetograms obtained by Solar Dynamics Observatory (SDO), we study the detailed process of three homologous confined flares in active-region (AR) NOAA 11861 on 2013 October 12. All of the three flares occurred at same location, with similar morphologies and comparable class. Through analyzing the evolution of magnetic field and flow field, we found an emergence of magnetic flux and a strong shearing motion between two opposite polarities near the following sunspot. The magnetic flux and the average transverse field strength exhibited a decrease before each eruption and reached the lowest point at the onset of each eruption. By calculating the shearing and the emergence energy in the photosphere, we found that the integral of energy injected from the photosphere, for a few hours, could provide enough energy for the flares. The reconnection between different loops was observed in H$\alpha$ images during the occurrence of each flare. These results suggest that the emerging magnetic flux and the shearing motion in the photosphere can inject the energy to the sheared magnetic loops and the energy finally was released via magnetic reconnection to power the solar flares.
\end{abstract}
\keywords{Sun: activity $-$ Sun: flares $-$ Sun: sunspots $-$ Sun: magnetic fields $-$ Sun: chromosphere $-$ Sun: energy}
\citep{} \cite{}
\section{Introduction}\label{sec:introduction}
Solar flares are thought to be one of the violent energetic events that are involving in the sudden release of magnetic energy stored in active region (AR), which can be transformed into the kinetic of particles and heating energy. Some flares are accompanied by coronal mass ejections (CMEs), which eject large amount of magnetic structures into interplanetary space \citep{1995A&A...304..585H} and are known as eruptive flares. However, some flares are not associated with CMEs, which are called as confined flares \citep{1992LNP...399....1S,2010ApJ...725L..38G,2012A&A...541A..49Z}. The eruptive flares tend to be large intensity and long duration event \citep{1989ApJ...344.1026K,2003SoPh..218..261A}. The investigations revealed that the eruptions are greatly determined by surrounding magnetic structures. Break-out model to explain eruptive events is that, the overlying magnetic loops should exist an opening process \citep{1999ApJ...510..485A,2003NewAR..47...53L,2006SSRv..123..251F}. The torus instability plays an important role in solar eruptions. {If the decay index of overlying magnetic field strength with height is lower than 1.5, the solar eruption will be confined \citep{2006PhRvL..96y5002K,2007ApJ...668.1232F}.} \cite{2013ApJ...778...70C} and \cite{2014ApJ...793L..28Y} found that overlying arcade loops can prevent filament being eruptive.

Studies on the trigger mechanisms of solar flares are quite important in predicting the occurrence of flares. The flares with a similar origin within the AR and common spatial and temporal characters can be implied as homologous flares \citep{2002ApJ...566L.117Z,2004ApJ...612..546S,2004ApJ...611..545G,2007SoPh..240..283G,2015ApJ...808L..24C,2016ApJ...829L...1L}. To explain the triggering of solar flares, several theories have been proposed by many researchers. In magnetohydrodynamics (MHD) models, solar flares can be triggered by MHD instabilities, e.g., kink instability \citep{1979SoPh...64..303H,2005ApJ...630L..97T}. As the studies of \cite{2018SSRv..214...46G}, the torus instability, as an ideal MHD instability, is responsible for the acceleration of the ejecta and considered as a driving mechanisms in solar eruptions\citep{2006PhRvL..96y5002K,2010ApJ...718.1388D}. Using data-driven simulation, \cite{2016ApJ...828...62J} suggest that eruption can be triggered by jet-like reconnection. Magnetic cancellation is also an major trigger mechanism for solar flares \citep{1989SoPh..121..197L,2011A&A...526A...2G}. Sigmoidal flux ropes can be built by photospheric flux cancellation and accumulated enough stress that make magnetic structure to erupt \citep{2012ApJ...759..105S}. Some researches suggest that the pre-flare activities may be a result of slow reconnection \citep{2001ApJ...547L..85K}. The pre-flare activities caused by magnetic reconnection play a crucial role in destabilizing the magnetic field, leading to a solar eruptive flare and associated large-scale phenomena \citep{2011ApJ...743..195J}. Reconnection of a strongly sheared field below the magnetic arcades can trigger solar eruptions, which correspond to the tether-cutting scenario \citep{2012ApJ...750...24M}.

Solar flares are also associated with the rapid rotation of sunspots \citep{2007ApJ...662L..35Z,2009RAA.....9..596Y,2009ApJ...706L..17L,2009SoPh..258..203M,2012ApJ...754...16Y,2013SoPh..286..453T}. The sunspot motions help in the building up of magnetic shear at many locations of the action region. The rotation direction of sunspots opposite to the differential rotation of the sun have higher X-class flare production \citep{2008MNRAS.391.1887Y}. Further study of rotating sunspots, \cite{2008MNRAS.391.1887Y} and \cite{2012ApJ...744...50J} ravel that active region's production rate of flares is related to the rotating sunspots. Recently, the investigations show that shearing motion and sunspots rotation play an import role in buildup of free energy and the formation of flux ropes \citep{2018ApJ...856...79Y}. Besides, the new emergence of magnetic flux can cause the destabilization of sheared fields and eventually leading to eruption \citep{1998SoPh..179..133C}.

As is well known, magnetic energy  consists of both the potential energy and free energy. The higher ratio of free energy to potential energy of active regions can make the field become less stable that the magnetic field will be capable of being triggered to explode when the ratio getting the crucial point \citep{2012ApJ...750...24M}. It has been thought that free energy stored in active regions can be as a result of twist or shear of magnetic field near the polarity inversion lines(PIL) \citep{2008ApJ...689.1433F,2012ApJ...750...24M}. A strong gradient of magnetic field with highly sheared coronal field likely produces solar flares \citep{1968SoPh....5..187M,1990ApJS...73..159H,2006ApJ...649..490W}. Some researches show that the free energy is initial from emergence magnetic field, and later the shearing flows build up much of it \citep{2012ApJ...761..105L}. The emergence of new flux at the onset of flare has sufficient energy to power flares and makes the magnetic field become a more sheared configuration \citep{2018A&A...612A.101V}. Meanwhile, fast rotation of sunspot and shearing motions can store sufficient energy to account for solar flares \citep{2009ApJ...704.1146K,2010ApJ...722.1539K}. By calculating the free energy of magnetic field, \cite{2015ApJ...804L..20A} found a decreasing of free energy after flares. Magnetic energy can be transported from the solar interior to the corona by the photospheric motions \citep{1984JFM...147..133B,1996SoPh..163..319W,2003SoPh..215..203D}. The horizontal motion transfers the magnetic energy from the photosphere to upper atmosphere, while the vertical motion carries the energy from below the photosphere\citep{2002ApJ...577..501K}. The magnetic energy flux transportation across the photosphere can be defined as follows \citep{2002ApJ...577..501K,2012ApJ...761..105L}:
\[\frac{\mathrm{d}E}{\mathrm{d}t}\Biggl\vert_{S}=\frac{1}{4\pi}\int_{S}B_{t}^{2}V_{\perp\,n}dS-\frac{1}{4\pi}\int_{S}(B_{t}\cdot\,V_{\perp\,t})B_{n}dS.\]
Where $B_t$ and $B_n$ refer to the horizontal and vertical components of magnetic field, $V_{\perp\,t}$ and $V_{\perp\,n}$ denote the horizontal and vertical components, with respect to the photosphere, of the plasma velocity perpendicular to the magnetic field. The first and the second terms are caused by the vertical motion through the photosphere and horizontal motion in the photosphere, respectively, which means the energy flux across photosphere generated by the emergence of twisted magnetic tube from solar interior and the shearing motion on surface.
\begin{figure}
\plotone{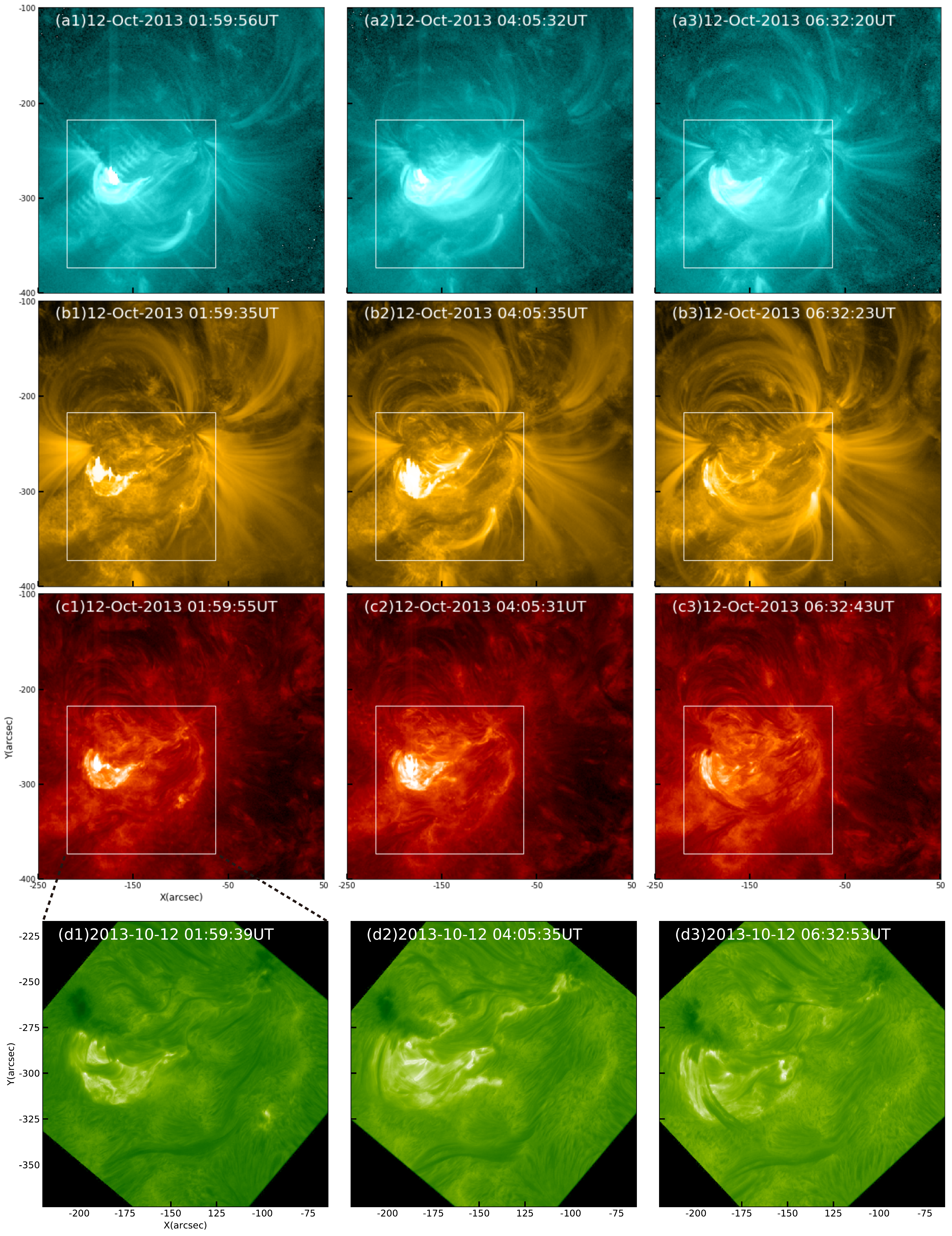}
\caption{Three confined flares observed by SDO and NVST. Panels (a1)-(c3) show the three confined eruptions in AR NOAA 11861 acquired at 131 \AA, 171 \AA, 304 \AA\ images, from upper row to lower row, the white box outlines the FOV of H$\alpha$ images. Panels (d1)-(d3) refer to the eruptions of three flares in the H$\alpha$ images. The three columns refer to the first, the second and the third eruptions, respectively. An animation with composite images is available in the online Journal. The animation includes the SDO 131 \AA, 171 \AA, 304 \AA\ and 1600 \AA\ images, and runs from $01:00$ to $07:00$ UT. \label{FIG1}}
\end{figure}

To study the energy origination of solar flares, a series of confined flares occurring in AR NOAA 11861 were investigated in detail in this paper. These events were first studied by \cite{2014ApJ...793L..28Y}. Their researches found that the overlying loops help to prevent the flares being eruptive and the emerging loops interact with pre-existing loops cause the reconnection that triggered the flares. In this paper, we present more detail research to show the evidence of flux emergence and shearing motion in photosphere. What is more, we calculate the energy flux across the photosphere, containing shearing and emergence energy, and try to find out that whether the energy of the flares can be accumulated by the emergence of flux and shearing motion in the photosphere. The detail observations are presented in Section \ref{sec:observations}. The results are shown in Section \ref{sec:results}. We discuss and make a conclusion in Section \ref{sec:conclusion}.

\section{Observations}\label{sec:observations}
The New Vacuum Solar Telescope (NVST) is one meter vacuum solar telescope that locates in the southwest of China, and aims to observe the fine structures in both the photosphere and the chromosphere \citep{2014RAA....14..705L}. It has three imaging channels, which consists of one chromospheric channel and two photospheric channels. The band for observing the chromosphere is H$\alpha$ (6562.8 \AA) with a bandwidth of 0.25 \AA. The NVST data that we adopted in this paper were H$\alpha$ center images, from 01:01:32 UT to 08:00:02 UT on 2013 October 12 with a cadence of 12 s and a field-of-view of (FOV) 152$^{\prime\prime}$$\times$152$^{\prime\prime}$ with a pixel size of 0.164$^{\prime\prime}$.
\begin{figure}
\plotone{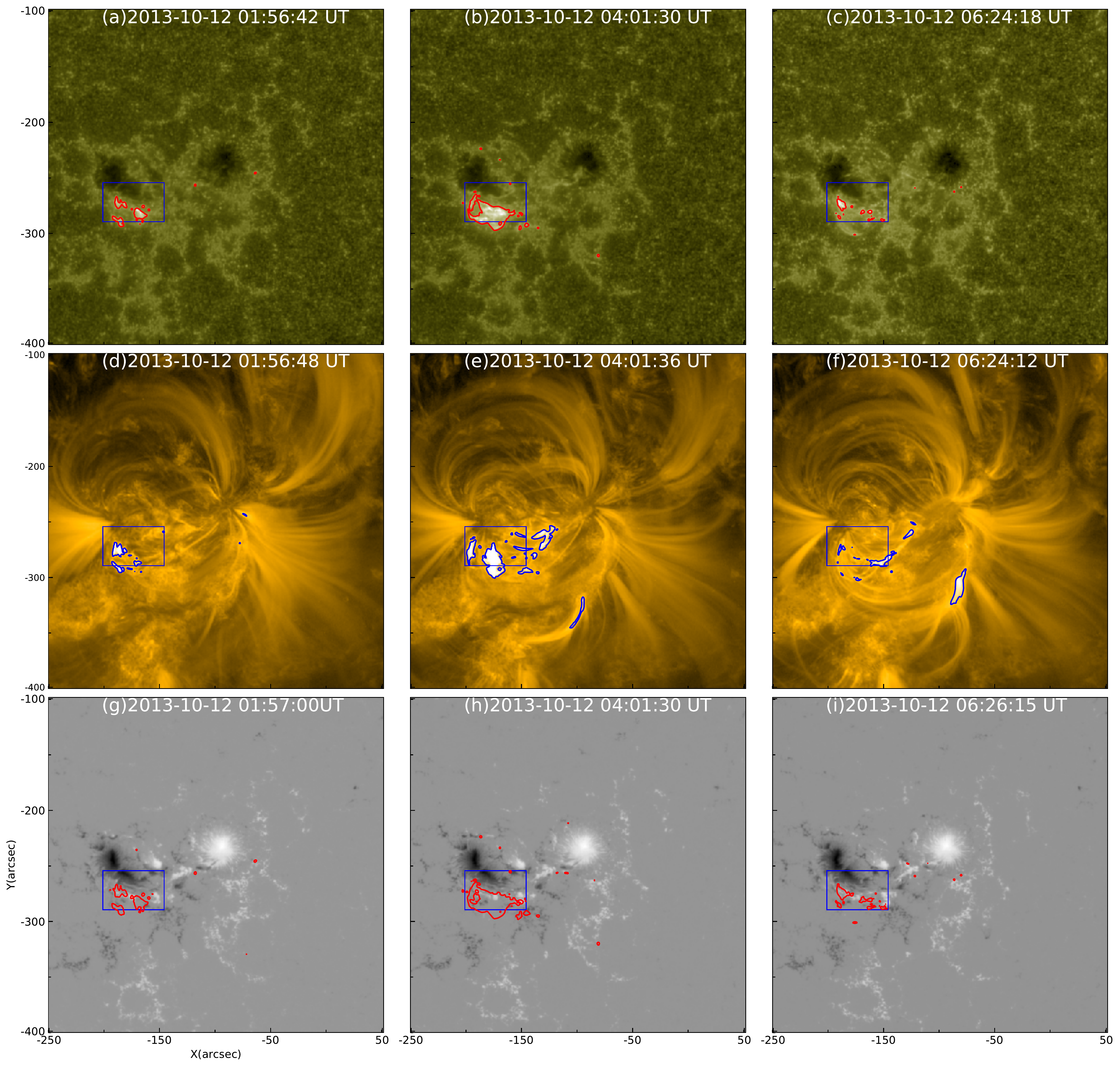}
\caption{Evolution of active region in 1600, 171 \AA\ images and LOS magnetograms. Panels (a)-(f) refer to the 1600, 171 images at the onset stage of the three confined flares. Panels (g)-(i) show the LOS magnetogram at the initial stage of the three eruptions. The white and the black colors refer to the positive and the negative magnetic polarities, respectively. {The red and the blue contours in panels (a)-(f) indicate the sites of brightening in 1600 \AA\ and 171 \AA\ images, respectively, and the red contours in 1600 \AA\ images are included on the LOS magnetograms.} The blue box denotes the field of view of Figures \ref{FIG3} and \ref{FIG6}. \label{FIG2}}
\end{figure}
Meanwhile, the simultaneous SDO/Atmospheric Imaging Assembly(AIA) \citep{2012SoPh..275...17L} multi-band images are also used. The 304, 171, 131, 1600 \AA\ images, with a pixel size of 0.6$^{\prime\prime}$, are employed to show the eruption process of the three confined flares. To identify the flares better, the Geostationary Operational Environmental Satellite (GOES) data are also used to show the variation of soft X-ray flux. According to GOES X-ray data, three flares occurred at around 01:58 UT, 04:05 UT, 06:32 UT, respectively. We set the reference time at 01:00 UT. The images from all of the three AIA channels were prepared to Level 1.5 and also co-aligned to the reference time by using the standard procedure in SSW. The LOS magnetograms with 45 s cadence, obtained by SDO/Helioseismic and Magnetic Imager (HMI) on board the SDO from 2013 October 11 to 12, were used to show the evolution of magnetic field. The vector magnetograms observed by HMI \citep{2012SoPh..275..229S,2014SoPh..289.3549B,2014SoPh..289.3531C,2014SoPh..289.3549B} are employed to show the evolution of vertical and horizontal magnetic fields. These magnetograms from the Space Weather HMI Active Region Patch (SHARP) series have a cadence of 12 minutes and a pixel scale of about 0.5$^{\prime\prime}$. The vector field data are derived by using the Very Fast Inversion of the Stokes Vector algorithm \citep{2011SoPh..273..267B}. Meanwhile, the minimum energy method \citep{1994SoPh..155..235M,2006SoPh..237..267M,2009SoPh..260...83L} is used to resolve the $180\,^{\circ}$ azimuthal ambiguity.
\begin{figure}
\plotone{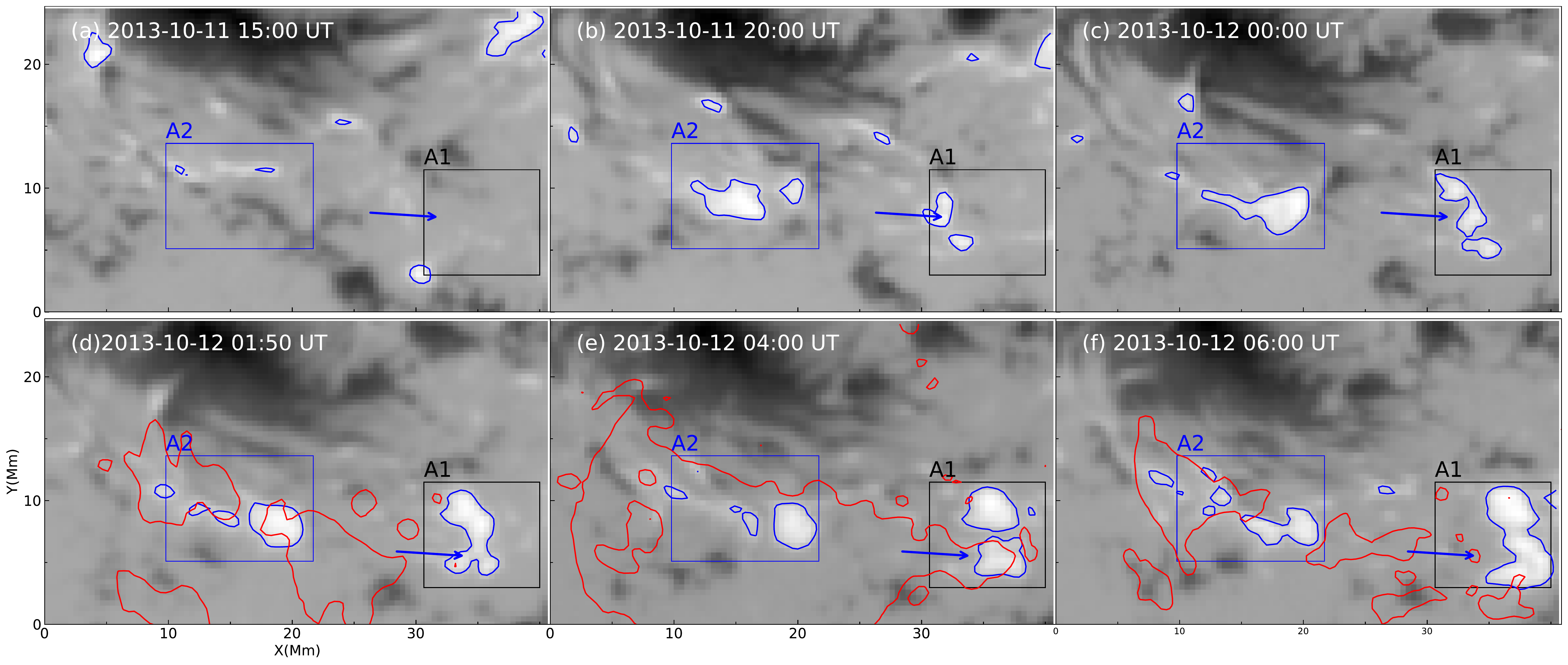}
\caption{Evolution of magnetic field in the LOS magnetograms, from 15:00 UT on 2013 October 11 to 06:00 UT on 2013 October 12. The blue arrows indicate the position of flux emergence. The blue curves are the contours of the positive field and the red curves refer to the brightening region of 1600 \AA\ images at pre-flare stage. The black box (A1) and the blue box (A2) refer to the calculated region of Figure \ref{FIG4}. An animation of the LOS magnetograms is available in the online Journal that begins at $12:00$ UT on 2013 October 11 and ends at $06:34$ UT on 2013 October 12. \label{FIG3}}
\end{figure}
\section{Results}\label{sec:results}
AR NOAA 11861 first appeared at the east limb of solar surface on 2013 October 8, and was closing to the central meridian on 2013 October 12 with a type of $\beta\gamma\delta$ sunspot group. Figure \ref{FIG1} shows the images of three eruptions at 131 \AA\ (panels (a1)-(a3)), 171 \AA\ images (panels (b1)-(b3)), and 304 \AA\ images (panels (c1)-(c3)), observed by SDO/AIA. The general appearances of flares in H$\alpha$ images are shown in panels (d1)-(d3). The three flares occurred at around 01:58 UT, 04:05 UT and 06:32 UT (see panels (d1)-(d3)), respectively. It's obvious that, these eruptions occurred at the same location, with similar morphologies and comparable class (C5.2, C4.9, and C2.0). A sequences of flares, occurring in a sustaining magnetic structure with common spatial and temporal characters can be implied as homologous flares and tend have similar trigger mechanisms \citep{2002ApJ...566L.117Z,2015ApJ...808L..24C}. During these process, dark materials were seen to be ejected into solar atmosphere and fell back to solar surface along a large arcade (see the online animation associated with Figure~\ref{FIG1}). In the previous studies of \cite{2014ApJ...793L..28Y}, the time-distance diagram shows the overlying loops lifted at first as the flare expanded, and then stopped at a certain height when the flare began to decay. Moreover, 171 \AA\ images show the coronal loops maintained their initial configuration without significant change. According to the coronagraph images of SOHO/LASCO C2, there were no signatures of successfully-ejected material associated with these flares. Therefore, these flares can be called homologous confined flares.
\begin{figure}
\plotone{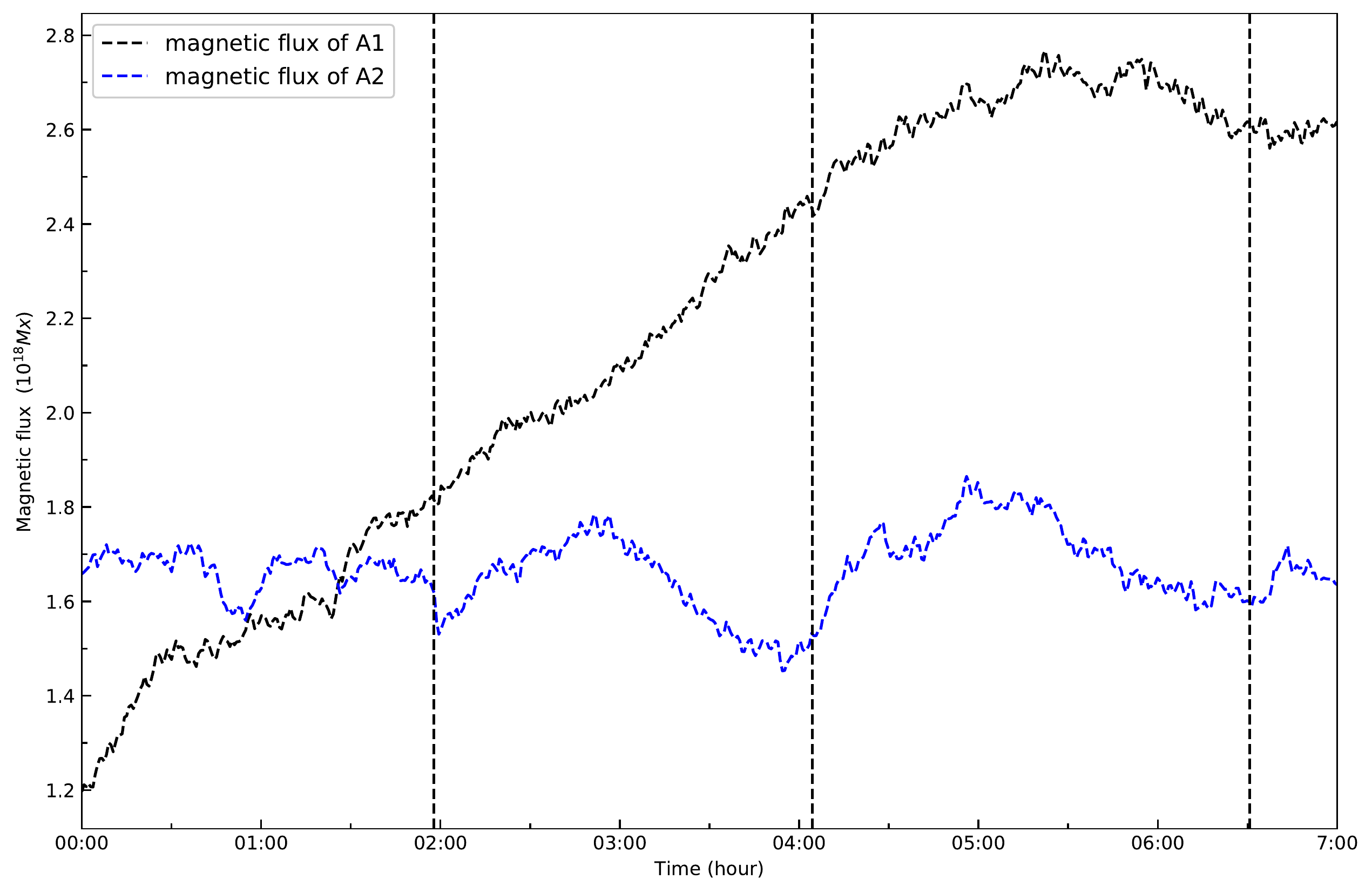}
\caption{Temporal profiles of positive magnetic flux from 00:00 UT to 7:00 UT on 2013 October 12. The black and the blue curves refer to the magnetic flux in the region outlined by black box (A1) and blue box (A2) in Figure \ref{FIG3}, respectively. The black vertical dashed lines refer to the peak time of three flares. \label{FIG4}}
\end{figure}

Previous studies show that, the sites of pre-flare brightening with respect to the location of flux rope/filament is essential to understand the triggering process of solar eruptions. Figure \ref{FIG2} shows the evolution of active region in 1600, 171 \AA\ images and LOS magnetograms. Panels (a)-(f) present the onset stage of the three eruptions acquired in 1600 \AA\ and 171 \AA. Panels (g)-(i) show the LOS magnetograms at the initial stage of three flares. The red and blue contours indicate the sites of pre-flare brightening in 1600 \AA\ and 171 \AA\ images respectively. The white and the black colors in the LOS magnetograms refer to the positive and the negative magnetic polarities, respectively. The 1600 \AA\ images indicate the locations of these flares. As shown in panels (a)-(c) and LOS magnetograms (panels (g)-(i)), all of the three flares located in the blue box. The position of the emerging flux is a crucial factor in the interaction of the background magnetic field and the resulting evolution \citep{2018ApJ...862..117D}. Flux emergence near a pre-existing, current-carrying magnetic structure is thought to be one possible mechanism for the initiation of eruption \citep{2000ApJ...545..524C}. To present the evolution of magnetic field in the blue box, shown in Figure \ref{FIG2}, detail structure of LOS {magnetograms} are presented in Figure \ref{FIG3}. The blue curves are the contours of positive {field} at 400 G (see panels (a)-(f)), and the blue arrows denote the position of flux emergence. The red curves refer to the brightening regions of three flares at the initial stage.

As is shown in Figure \ref{FIG3}, the areas of positive field in black box (A1), outlined by the blue curves and denoted by the blue arrow, were growing with time. There was almost not positive polarity in A2, as shown in panel (a). Next, it emerged rapidly (see panel (b)). This suggests that flux emergence existed within A1 and A2. There were several companied C-class flares occurred, with the emergence of flux in A1 and A2 from 15:00 UT on October 11 to 00:46 UT on October 12. It's no doubt that the flux emergence plays an critical role in the occurrence of these flares. The temporal profiles of positive magnetic flux, from 00:00 UT to 07:00 UT on 2013 October 12, are shown in Figure \ref{FIG4}. Black and blue curves refer to the positive magnetic flux in the area outlined by the black box (A1) and the blue box (A2) in Figure \ref{FIG3}, respectively. It's obvious that, positive magnetic flux in A1 increased from 00:00 UT to 06:00 UT. We found some interesting results that, the positive magnetic flux in A2 decreased before each flare and reach the lowest point at the onset of eruption and increased after each flare. This phenomenon implies that there may exist a process of flux cancelation within A2 caused by magnetic reconnection.
\begin{figure}
\plotone{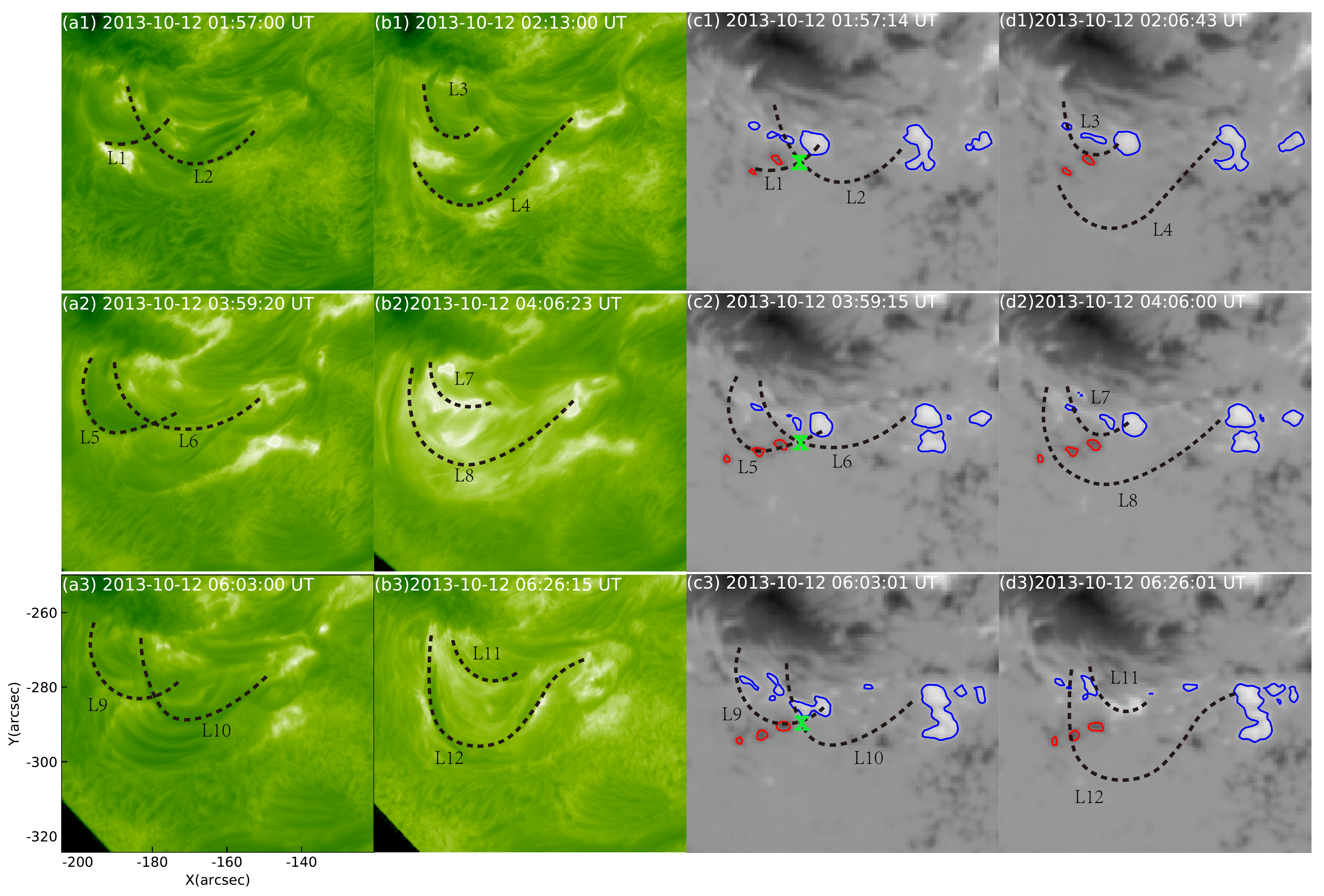}
\caption{Topology evolution acquired at H$\alpha$ images. Panels (a1)-(a3) and (b1)-(b3) display the process of the first, the second and the third flares obtained from H$\alpha$ data, respectively. Panels (c1)-(c3) and (d1)-(d3) refer to the evolution of LOS magnetograms, with the positive field outlined by blue curves and negative field by red. Black curves L1-L12 represent the visible loops identified in H$\alpha$ images. The green cross symbols in panel(c1)-(c3) mark the site where the reconnection occurred. An animation of the H$\alpha$ images is available in the online Journal. The animation runs from $01:32$ to $08:00$ UT and is annotated with the locations of the 12 loops and the reconnection. \label{FIG5}}
\end{figure}

To study the reconnection process, we use the H$\alpha$ images to get the fine structure of loops and its evolution process. As the previous studies of \cite{2014ApJ...793L..28Y}, magnetic reconnection took place between different loops, according to the observations of NVST. Figure \ref{FIG5} displays the evolution process of the three flares both in H$\alpha$ images and LOS magnetograms. At the pre-flare stage (panels (a1)-(a3)), many arch-shaped H$\alpha$ fibrils can be clearly observed, as delineated with curves L1, L2 (panel (a1)) and L5, L6 (panel (a2)) and L9, L10 (panel (a3)). A few minutes after the onset of solar flares, new loops L3, L4, L7, L8, L11 and L12 were formed (see panels (b1)-(b3)). As shown in LOS magnetograms (panels (c1)-(c3) and (d1)-(d4)), the arch-shaped H$\alpha$ fibrils, delineated in black curves, connect the positive polarity and the nearby negative fields. The magnetic reconnection took place between loops L1 and L2 at the site marked by the green cross symbol (see panel (c1)). Similar process took place between loops L5 and L6 (see panel (c2)) and between L9 and L10 (see panel (c3)). Due to the reconnection, loops L3, L4, L7, L8, L11, L12 were newly formed, as indicated by the black curves in panels (d1)-(d3).
\begin{figure}
\plotone{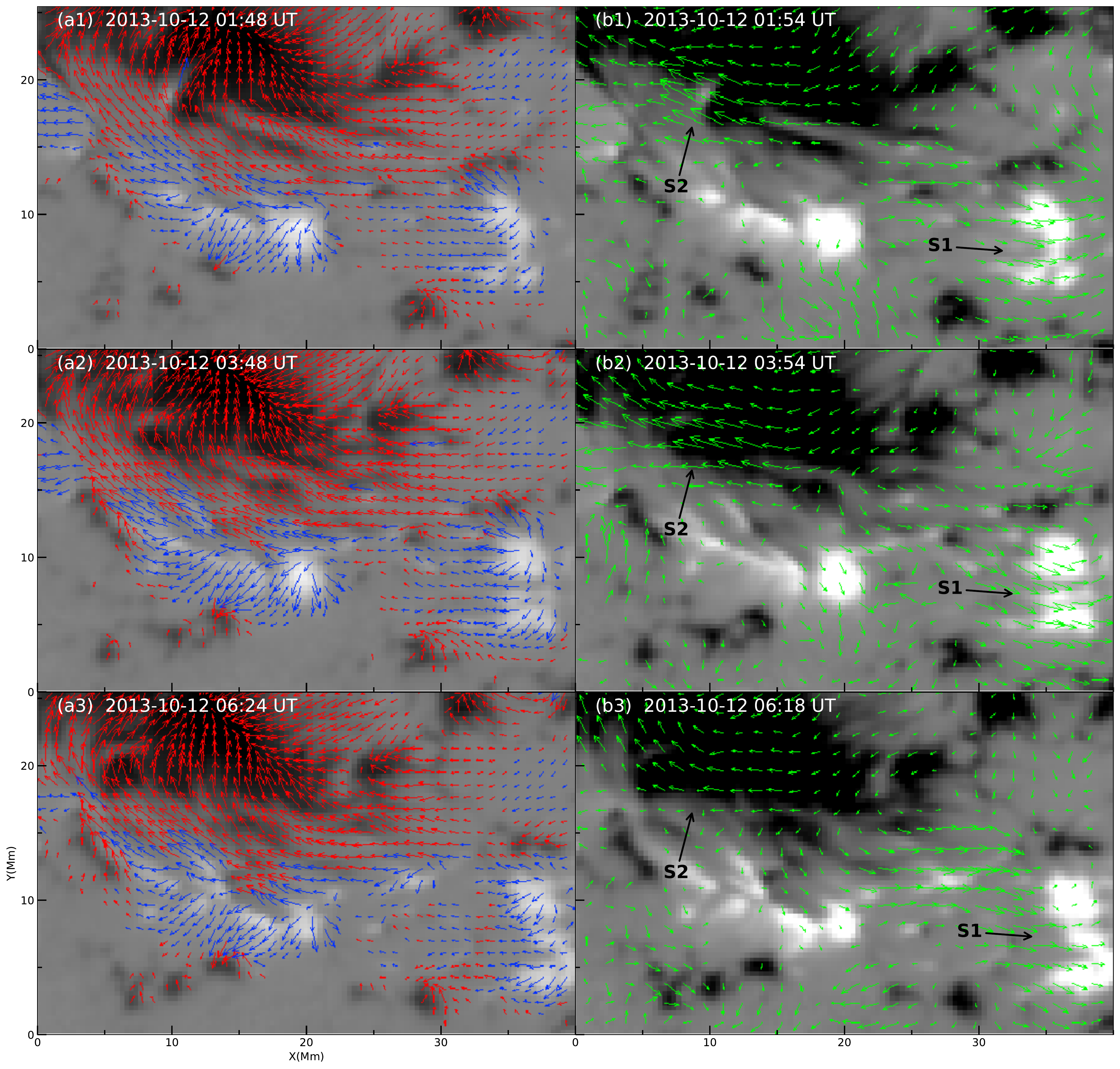}
\caption{Evolution of vector magnetograms and velocity maps, as same as the field-of-view of figure \ref{FIG3}. Panels (a1)-(b3): The vector magnetograms of our focus area, observed by HMI/SDO, with the positive field in white and negative field in black. The red and the blue arrows, in the left column, refer to the transverse magnetic fields of negative polarity and positive polarity, respectively. The green arrows in panels (b1)-(b3) refer to tangential velocity calculated by using the DAVE method. S1 and S2 denote positive polarity and negative polarity, respectively. \label{FIG6}}
\end{figure}
\begin{figure}
\plotone{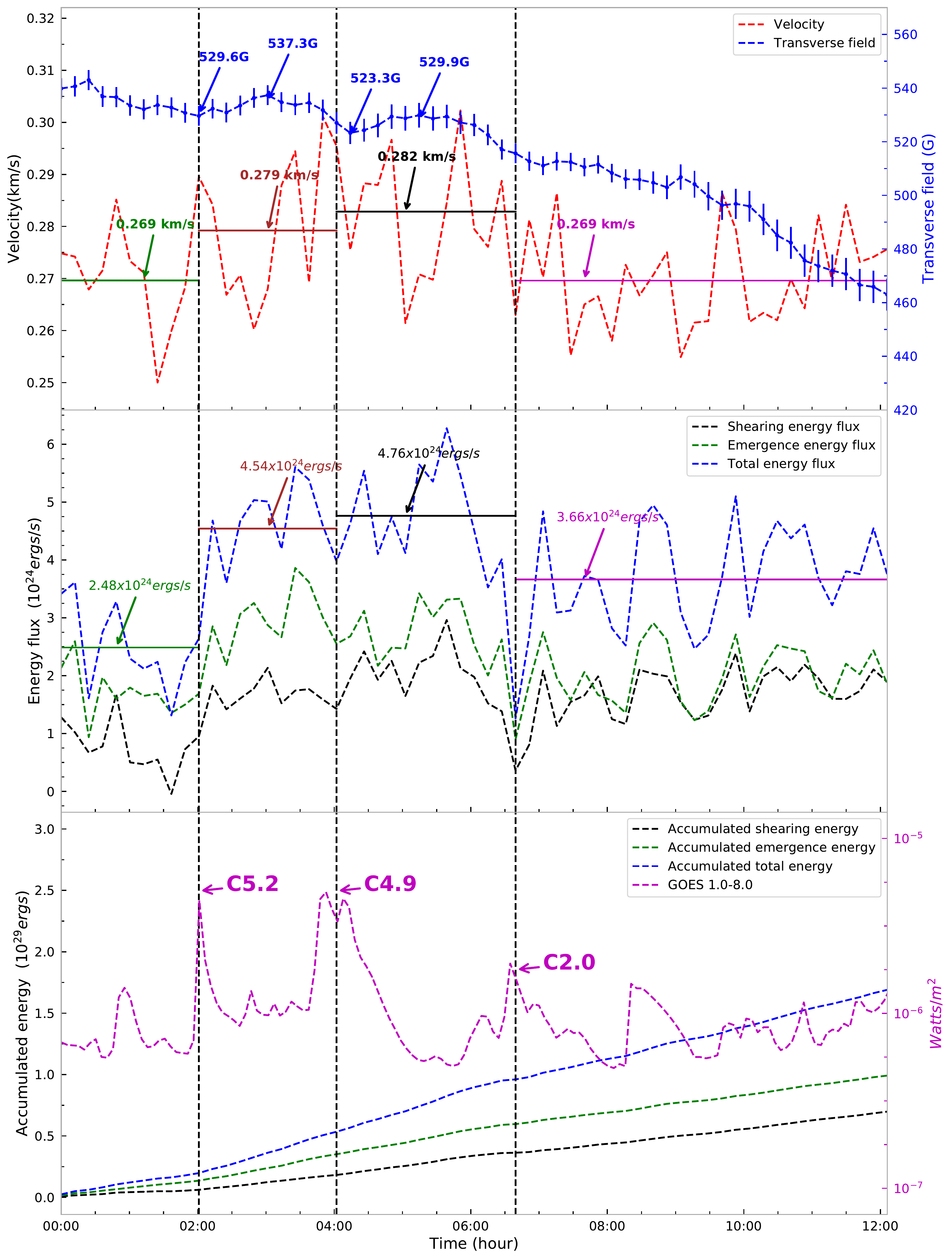}
\caption{Top: Temporal profiles of average flow velocity (red curve) and average transverse field strength (blue curve). The green, brown, black and pink horizontal lines refer to the average velocity from 00:00 UT to 02:00 UT, 02:00 UT to 04:05 UT, 04:05 UT to 06:35 UT and 06:35 UT to 12:05 UT, respectively. Middle: Temporal profiles of energy fluxes, the black, the green and the blue curve indicate the shearing energy flux, the emergence energy flux and total energy flux, respectively. The green, brown, black and pink horizontal lines refer to the average total energy flux from 00:00 UT to 02:00 UT, 02:00 UT to 04:05 UT, 04:05 UT to 06:35 UT and 06:35 UT to 12:05 UT, respectively. Bottom: The integral of energy flux over time, with shearing energy in black and emergence energy in green and the total energy in blue. The pink line indicates the profiles of GOES x-ray 1-8 \AA\ from 00:00 UT to 12:05 UT on 2013 October 12. The black vertical dashed lines refer to the peak time of three flares. \label{FIG7}}
\end{figure}

The evolution of vector magnetograms of our focus areas, outlined by blue box in Figure \ref{FIG2}, are shown in Figure \ref{FIG6}. The red and the blue arrows in left panels (a1), (a2) and (a3) are referred to transverse magnetic fields of positive polarity and negative polarity, respectively. The magnetic fields between positive polarity and negative polarity show a strong shear {angle} along polarity inversion line (PIL), which means that the magnetic fields had a strong departure from potential field before each eruption. The green arrows in right panels (b1), (b2) and (b3) are referred to the direction of transverse flow field calculated by the differential affine velocity estimator (DAVE) method. It's obvious that the positive polarity (S1) and the negative polarity (S2) were moving away from each other, which denotes that there existed a shearing motion between S1 and S2.

To further understand the evolution of magnetic field, we calculate the average transverse field strength and the average flow velocity of the region outlined by the blue box that shown in Figure \ref{FIG2}. We also calculate the energy flux across the photosphere in the same region outlined by the blue box shown in Figure \ref{FIG2}, to determine how much accumulated energy of the eruptions are injected. The top panel of Figure \ref{FIG7} refers to the temporal profiles of average flow velocity (red curve) and average transverse field strength (blue curve). At the peak time of the first flare, 01:58 UT, the average transverse field strength is 529.6 G. It became 537.3 G at 03:00 UT, 523.3 G at 04:12 UT, and 529.9 G at 05:12 UT. One can see that the average transverse field strength decreased before each flare and had a small increasing after eruption, like the temporal evolution of magnetic flux of A2. But the overall profile of the average transverse field strength is decreasing during our calculate duration. Due to the noisy data of average flow velocity, we make an average value within a period of time. The green solid lines refer to the average velocity from 00:00 UT to 02:00 UT. The brown solid lines refer to the average velocity from 02:00 UT to 04:05 UT. The black solid lines refer to the average velocity from 04:00 UT to 06:35 UT. The pink solid lines refer to the average velocity from 06:35 UT to 12:05 UT. The average velocity from 02:00 UT to 04:05 UT and 04:35 UT to 06:35 UT are about 0.279 $km/s$ and 0.282 $km/s$, respectively, while the average velocities from 00:00 UT to 02:00 UT and 06:35 UT to 12:05 UT are all about 0.269 $km/s$. It turns out that the flow velocity is a little faster during the period of three flares, from 02:00 UT to 06:35 UT. The shearing energy flux (black), emergence energy flux (green) and total energy flux (blue) are shown in middle panel. {As with the average flow velocity}, we make an average value within a period of time. The green solid lines refer to the average total energy flux from 00:00 UT to 02:00 UT. The brown solid lines refer to the average total energy flux from 02:00 UT to 04:05 UT. The black solid lines refer to the average total energy flux from 04:00 UT to 06:35 UT. The pink solid lines refer to the average total energy flux from 06:35 UT to 12:05 UT. The average total energy flux is 2.48 $\times$ 10$^{24}$ ergs/s from 00:00 UT to 02:00 UT, 4.54 $\times$ 10$^{24}$ ergs/s from 02:00 UT to 04:05, 4.76 $\times$ 10$^{24}$ ergs/s from 04:00 UT to 06:35 UT, and 3.66 $\times$ 10$^{24}$ ergs/s from 06:35 UT to 12:05 UT. As the evolution of the average flow velocity, the average total energy flux is more larger during the period of three flares, from 02:00 UT to 06:35 UT. The integral of the energy flux over time are show in the bottom panel, with shearing energy in black and emergence energy in green. The blue curve is the summation of the two terms. GOES X-ray flux 1.0-8.0 \AA\ (pink), shown in bottom panel, are used to determine the eruption time and class of the three flares. Both of the energy fluxes evolved consistently in phase and the emergence energy flux was dominant, contributing much to the total energy. Total energy accumulated from 02:00 UT to 04:05 UT is about 0.34 $\times$ 10$^{29}$ ergs, which is comparable to the energy accumulated from 04:05 UT to 06:35 UT (0.43 $\times$ 10$^{29}$ ergs). The integral of energy is around 10$^{29}$ ergs and comparable with a C-class flare.
\section{Conclusion and discussion}\label{sec:conclusion}
Active region NOAA 11861 produced three confined flares from 02:00 UT to 06:35 UT on 2014 October 12. All of the three flares occurred at a same location, had a similar morphologies, and a comparable classes (C5.2, C4.9, and C2.0). Thus, these flares can be implied as homologous flares, which may have an analogous trigger mechanisms.

During the evolution of active region, the magnetic emergence and cancellation were found before and during three flares. According to the temporal profile of positive magnetic flux, a continue process of flux emergence happened in region A1. In the studies of \cite{2002ApJ...566L.117Z}, the continuous emergence of moving magnetic features trigger the homologous flares. Moreover, the positive magnetic flux of A2 decreased before each of the flares and reached the lowest point around the onset of the eruptions and then increased. Previous investigations found that magnetic flux tends to decrease during flares rather than increase \citep{2013SoPh..283..429B}. Our results found that magnetic flux decreased before these flares and became the smallest around the onset of the eruptions. By carefully examining the evolution of magnetic fields, it is found that flux emergence first occurred before the flares and then flux cancelation is followed in the region A2. Many models have added the flux emergence into a pre-existing magnetic field configuration to explain the eruptions \citep{2008ApJ...689L.157Z,2009A&A...507..441Z,2012SoPh..280..389J,2012ApJ...760...31K,2012NatPh...8..845R}. It suggests that the emergence of new magnetic flux plays an important role in the onset of flares. Magnetic reconnection occurs between the new magnetic loops that emergence from below photosphere and the pre-existing overlying field. Magnetic reconnection in destabilizing the magnetic configuration is considered as a possible mechanism to trigger solar eruptions \citep{2008AnGeo..26.3089V}. We can deduce that the existence of flux emergence may be crucial in triggering the sequential confined flares. In these events, it implies that the reconnection took place between a set of loops, according to the H$\alpha$ images observed by NVST \citep{2014ApJ...793L..28Y}.

As revealed in previous observations, horizontal shear flows are responsible for magnetic stress and driver for magnetic reconnection \citep{2006SoPh..239..317Y}. The transverse magnetic fields and the velocity maps, calculated by using the DAVE method, show that there had a strong shear angle along PIL and shear flow between the polarities S1 and S2 before each of eruption. The temporal evolution of velocity maps show that there was a faster velocity during the period of these flares, from 02:00 UT to 06:35UT. Analysis shows that average transverse field strength decreased before the onset of each flare and exhibited a small increasing after each flare. The decreasing of the average transverse field strength and magnetic flux of A2 calculated from LOS magnetograms before the flares suggest that the magnetic cancellation, as a result of reconnection, could be going on before the flares.

Since the shearing motion and flux emergence can drive magnetic reconnection, the evidence of magnetic reconnection between different loops was also found in H$\alpha$ images. The three flares were all triggered by magnetic reconnection between two groups of loops. By calculating the free energy associated with flares, a significant drop of free energy was found after flare\citep{2009ApJ...696...84J}. The energy across the photosphere was calculated in previous studies. \cite{2012ApJ...761..105L} found the source of energy injection is the areas with strong up-flows in the surrounding of leading sunspot. \cite{2018ApJ...865..139B} found the energy flux across the sunspot was negative during the flare and positive before the flare. However, whether the energy across the photosphere can support to a specific flare is not sure. We calculate the energy flux in the area, where the three flares occurred. As the studies of \cite{2012ApJ...761..105L}, the shearing and the emergence fluxes evolved consistently in phase, while the emergence flux contribute much of the energy. The total energy flux seems much larger before the end of last flare. Furthermore, the energy accumulated by shearing and vertical motion in the photosphere, from 00:00 UT to 06:35 UT, was enough to support the energy of eruptions. Note that, the accumulated energy could not totally been released during the flares. Moreover, since free energy may be already stored in magnetic field before the occurrence of the first flare. Therefore the energy accumulated by shearing and vertical motion in the photosphere may provide partial energy of these flares. Meanwhile, the shearing motion and flux emergence drive the magnetic reconnection between the different loops to destabilize the topology of magnetic structure and finally trigged these eruptions.

We would like to thank the NVST, SDO/ AIA, and SDO/ HMI teams for the high-cadence data support. This work is sponsored by the National Science Foundation of China (NSFC) under the grant numbers 11873087, 11603071,11503080, 11633008, by the Youth Innovation Promotion Association CAS (No.2011056) , by the Yunnan Science Foundation of China under number 2018FA001, by Project Sup-ported by the Specialized Research Fund for State Key Laboratories and by the grant associated with project of the Group for Innovation of Yunnan Province.

\end{document}